\documentclass[journal]{IEEEtran}

\ifCLASSINFOpdf

\else

\fi

\usepackage{soul}
\usepackage{algorithm}
\usepackage[noend]{algpseudocode}
\usepackage{amsmath}
\usepackage{graphicx}
\usepackage{amssymb}
\usepackage{color}
\usepackage{graphicx}
\usepackage{subfig}
\usepackage{caption}
\usepackage{cite}
\usepackage{hyperref}
\let\oldref\ref
\renewcommand{\ref}[1]{(\oldref{#1})}

\begin{document}

\title{Interpretable multimodal fusion networks reveal mechanisms of brain cognition}
\author{Wenxing~Hu, Xianghe~Meng, Yuntong~Bai, Aiying~Zhang, Biao~Cai, Gemeng~Zhang, Tony~W.~Wilson, Julia~M.~Stephen,  Vince~D.~Calhoun,~\IEEEmembership{Fellow,~IEEE,}
        and~Yu-Ping~Wang,~\IEEEmembership{Senior Member,~IEEE}
\thanks{YP. Wang* is with the Biomedical Engineering Department, Tulane University, New Orleans, LA 70118. (corresponding author e-mail: wyp@tulane.edu).}
\thanks{W. Hu, Y. Bai, A. Zhang, B. Cai, G. Zhang are with the Biomedical Engineering Department, Tulane University, New Orleans, LA 70118.}
\thanks{X. Meng is with the Center of System Biology, Data Information and Reproductive Health, School of Basic Medical Science, Central South University, Changsha, Hunan, 410008, China. (e-mail: xhmeng2020@csu.edu.cn).}
\thanks{J. M. Stephen is with the Mind Research Network,
Albuquerque, NM 87106.}
\thanks{T. W. Wilson is with the Department of Neurological Sciences, University
of Nebraska Medical Center, Omaha, NE 68198.}
\thanks{V. D. Calhoun is with the Tri-institutional Center for Translational Research in Neuroimaging and Data Science (TReNDS), Georgia State University, Georgia Institute of Technology, Emory University, Atlanta, GA 30030. (e-mail: vcalhoun@gsu.edu).}
\thanks{Manuscript received **** **, ****; revised **** **, ****.}}

\markboth{Journal of \LaTeX\ Class Files,~Vol.~**, No.~*, ****~****}%
{Shell \MakeLowercase{\textit{et al.}}: Bare Demo of IEEEtran.cls for IEEE Journals}

\maketitle
\begin{abstract}
Multimodal fusion benefits disease diagnosis by providing a more comprehensive perspective. Developing algorithms is challenging due to data heterogeneity and the complex within- and between-modality associations. Deep-network-based data-fusion models have been developed to capture the complex associations and the performance in diagnosis has been improved accordingly. Moving beyond diagnosis prediction, evaluation of disease mechanisms is critically important for biomedical research. Deep-network-based data-fusion models, however, are difficult to interpret, bringing about difficulties for studying biological mechanisms. In this work, we develop an interpretable multimodal fusion model, namely gCAM-CCL, which can perform automated diagnosis and result interpretation simultaneously. The gCAM-CCL model can generate interpretable activation maps, which quantify pixel-level contributions of the input features. This is achieved by combining intermediate feature maps using gradient-based weights. Moreover, the estimated activation maps are class-specific, and the captured cross-data associations are interest/label related, which further facilitates class-specific analysis and biological mechanism analysis.
We validate the gCAM-CCL model on a brain imaging-genetic study, and show gCAM-CCL's performed well for both classification and mechanism analysis. Mechanism analysis suggests that during task-fMRI scans, several object recognition related regions of interests (ROIs) are first activated and then several downstream encoding ROIs get involved. Results also suggest that the higher cognition performing group may have stronger neurotransmission signaling while the lower cognition performing group may have problem in brain/neuron development, resulting from genetic variations.
\end{abstract}

\IEEEpeerreviewmaketitle

\section{Introduction}

\IEEEPARstart{R}{ecently}, there is increasing recognition that multimodal imaging data fusion can exploit the complementary information across different data, leading to better performance in terms of diagnosis and the analysis of mechanisms \cite{sui2020neuroimaging}. Conventional multimodal fusion is often focused on matrix decomposition approaches. Among these methods, canonical correlation analysis (CCA) \cite{cca} has been widely used to integrate multimodal data by detecting linear cross-data correlations. However, CCA fails when data have complex nonlinear interactions. To capture complex cross-data associations, deep neural network (DNN) based models, e.g., deep CCA \cite{deepcca}, have been developed which employ deep network to extract high-level cross-data associations. These methods can lead to improved performance in terms of prediction/diagnosis \cite{deepcca, wang2015deep}.

Beyond diagnosis, it is also important to uncover hidden disease mechanisms. This requires the data analysis model to be interpretable, i.e., with explicit and interpretable data representations. However, DNN is composed of a large number of layers and each layer consists of several nonlinear transforms/operations, e.g., nonlinear activation and convolution, resulting in difficulties in interpreting its data representations. Moreover, the captured cross-data associations are not guaranteed to be relevant to the variable of interest, e.g., disease. Instead, the associations may result from interest-irrelevant signals, e.g., noise and background. Therefore, it is not clear how to use the captured associations for disease mechanism analysis.

To address these issues, we develop an interpretable DNN based multimodal fusion model, Grad-CAM guided convolutional collaborative learning (gCAM-CCL), which can perform automated diagnosis and result interpretation simultaneously. The gCAM-CCL model can generate interpretable activation maps indicating pixel-wise contributions of the inputs, enabling automated result interpretation. Moreover, the activation maps are class-specific, which can further promote class-difference analysis and biological mechanism analysis. In addition, the cross-data associations captured by gCAM-CCL are interest-related, e.g., disease-related. This is achieved by feeding the network representations to a collaborative layer\cite{hu2019deep} which considers both cross-data interactions and the fitting to traits.

The rest of the paper is organized as follows. Section II describes the limitations of several existing multimodal fusion methods and how the proposed model addresses the limitations. Data collection and preprocessing procedures as well as experiments and results of applying gCAM-CCL to imaging genetic study can be found in Section III. A brief discussion was given in Section IV.

\section{Method}
\subsection{Multimodal data fusion: analyzing cross-data association}
Classical multimodal data fusion methods are often focused on cross-data matrix factorization. Among them, canonical correlation analysis (CCA) \cite{cca} has been widely used in multi-view/omics studies \cite{hu2017adaptive, lin2014correspondence}. CCA aims to find the most correlated variable pairs, i.e., canonical variables, and further association analysis can be performed accordingly.

Specifically, given two data matrices
$X_1 \in \mathbb{R} ^{n\times r}, X_2 \in \mathbb{R} ^{n\times s}$ ($n$ represents sample/subject size, and $r,s$ represents the feature/variable sizes in two data sets), CCA seeks two optimal loading vectors $u_1\in \mathbb{R} ^{r\times 1}$ and $u_2\in \mathbb{R} ^{s\times 1}$ which maximize the Pearson correlation $corr(X_1u_1, X_2u_2)$, as in Eq. \ref{eq:cca_vec}.
\begin{align}
& (u_1^*,u_2^*) = \mathop{\mbox{argmax}}\limits_{u_1,u_2} u_1'\Sigma_{12}u_2 \label{eq:cca_vec}\\
& \text{subject to } u_1'\Sigma_{11}u_1 = 1 ,\; u_2'\Sigma_{22}u_2 = 1 \notag
\end{align}
where $u_1 \in \mathbb{R} ^{r\times 1}, u_2 \in \mathbb{R} ^{s\times 1}, \; \Sigma_{ij}:=X_{i}'X_j, \; i,j = 1,2.$\\
Solving optimization Eq. \ref{eq:cca_vec} will yield the most correlated canonical variable pair, i.e., $X_1u_1$ and $X_2u_2$. More correlated canonical variable pairs (with lower correlations) can be obtained subsequently by solving the extended optimization problem, as formulated in Eq. \ref{eq:cca_mat}.
\begin{align}
& (U_1^*, U_2^*) = \mathop{\mbox{argmax}}\limits_{U_1,U_2} \text{Trace}\big(  U_1'\Sigma_{12}U_2\big) \label{eq:cca_mat}\\
& \text{subject to }  U_1'\Sigma_{11}U_1 = U_2'\Sigma_{22}U_2 = \mathbf{I}_n \notag
\end{align}
where $U_1 \in \mathbb{R} ^{r\times k}, U_2 \in \mathbb{R} ^{s\times k}, k=$ min(rank($X_1$), rank($X_2$)).

CCA captures only linear associations and therefore it requires that different data/views follow the same distribution. However, different modality data, e.g., fMRI imaging and genetic data, may follow different distributions and have different data structures. As a result, CCA fails to detect the association between heterogeneous data-sets. To address this problem, Deep CCA (DCCA) was proposed by Andrew et al. \cite{deepcca} to detect more complicated correlations. DCCA introduces a deep network representation before applying CCA framework. Unlike linear CCA, which seeks the optimal canonical vectors $U_1,U_2$, DCCA seeks the optimal network representation $f_1(X_1),\ f_2(X_2)$, as shown in Eq. \ref{deepcca_eqn}.
\begin{align}
& (f_1^{*},f_2^{*}) = \mathop{\mbox{argmax}}\limits_{f_1,f_2} \big\{ \mathop{\mbox{max}}\limits_{U_1,U_2} \frac{U_1'f_1'(X_1)f_2(X_2)U_2}{\|f_1(X_1)U_1 \|_2 \|f_2(X_2)U_2 \|_2} \big\} \label{deepcca_eqn}
\end{align}
where $f_1, f_2$ are two deep networks.\\
The introduction of deep network representation leads to a more flexible ability to detect both linear and nonlinear correlations. According to experiments on both speech data and handwritten digits data \cite{deepcca}, DCCA's representation was more effective in finding correlations compared to other methods, e.g., linear CCA, and kernel CCA. Despite DCCA's superior performance, the detected associations are not guaranteed to be  relevant to any phenotype of interest, e.g., disease. Instead, the detected associations, may be caused by irrelevant signals, e.g., background and noise. As a result, the use of detected associations is challenging for further disease mechanism analysis.

\subsection{Deep collaborative learning (DCL): phenotype-related cross-data association}
To address the limitations of DCCA, we proposed a multimodal fusion model, deep collaborative learning (DCL) \cite{hu2019deep}, which can capture phenotype-related cross-data associations by enforcing additional fitting to phenotype label, as formulated in Eq. \ref{cobj1-1}.
\begin{align}
(Z_1^*,Z_2^*) = & \mathop{\mbox{argmax}}\limits_{Z_1,Z_2} \{ \mathop{\mbox{max}}\limits_{U_1,U_2}\text{Trace}(U_1'Z_1'Z_2U_2) -
 \label{cobj1-1} \\
                                & \mathop{\mbox{min}}\limits_{\beta_1} \Vert Y - Z_1\beta_1 \Vert _2^2 - \mathop{\mbox{min}} \limits_{\beta_2} \Vert Y - Z_2\beta_2 \Vert _2^2  \} \notag \\
                = & \mathop{\mbox{argmax}}\limits_{Z_1,Z_2} \{
                 \Vert \Sigma _{11}^{-\frac{1}{2}} \Sigma _{12} \Sigma _{22}^{-\frac{1}{2}} \Vert _{tr} \notag \\
                 & - \Vert Y - Z_1(Z_1'Z_1)^{-1}Z_1'Y \Vert _2^2 \notag \\
                          & - \Vert Y - Z_2(Z_2'Z_2)^{-1}Z_2'Y \Vert _2^2  \} \notag \\
              = & \mathop{\mbox{argmax}}\limits_{Z_1,Z_2} F(Z_1,Z_2) \notag
\end{align}
where $U_1,\; U_2$ subject to  $U_1'\Sigma_{11}U_1=U_2'\Sigma_{22}U_2 = \mathbf{I}$; $\Vert A \Vert _{tr} := \text{Trace} (\sqrt{A'A}) = \Sigma \sigma_i $; $Z_1 = f_1(X_1) \in \mathbb{R} ^{n\times p},\; Z_2 = f_2(X_2) \in \mathbb{R} ^{n\times q}$,
$f_1,\ f_2$ represent two deep networks; $Y \in \mathbb{R} ^{n\times 1}$ represents phenotype or label data. 

As shown in Eq. \ref{cobj1-1}, DCL seeks the optimal network representation $Z_1 = f_1(X_1) ,Z_2=f_2(X_2)$ to maximize cross-data correlations. Compared to DCCA, DCL's representation retains label related information which guarantees label/phenotype related associations. In this way, further analysis of disease mechanisms can be performed and better classification performance can be achieved, according to the work described in \cite{hu2019deep}. Moreover, DCL relaxes the requirement that projections $u_1$ and $u_2$ have to be in the same direction. This leads to a better representation of both phenotypical information and cross-data correlation in a more effective manner.

With the ability to capture both cross-data associations and trait-related signals, DCL can exploit complementary information from multimodal data, as demonstrated in a brain imaging study \cite{hu2019deep}. However, DCL uses deep networks to extract high-level features, which are difficult to interpret, result in obstacles for identifying significant features/biomarkers. As a result, DCL can only be used for classification/diagnosis rather than exploring disease mechanisms, and consequently the medical impact of its applications is limited.
\begin{figure*}[h]
\begin{center}
\includegraphics[width=0.92\textwidth]{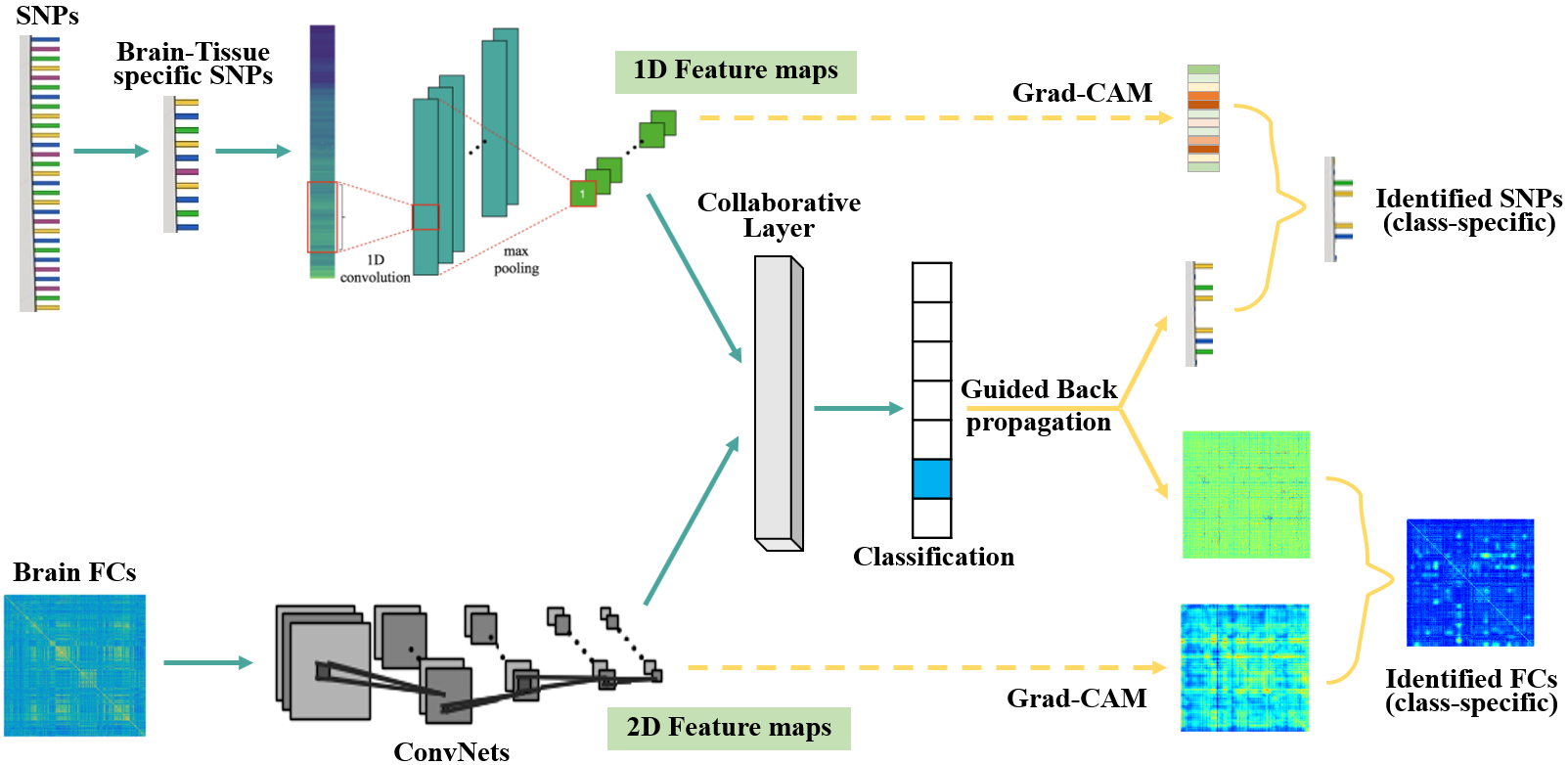}
\end{center}
\caption{The work-flow of convolutional collaborative learning (CCL), an end-to-end model for automated classification and interpretation for multimodal fusion. Genetic data is fed into a ConvNet and then flattened to a fully connected (FC) layer. Brain functional connectivity (FC) data is fed into a deep network. A collaborative learning layer fuses the two deep networks and passes two composite gradients mutually during the back-propagation process.} \label{fig:gcam-ccl_net}
\end{figure*}

\subsection{Deep Network Interpretation: CAM based methods}
Both DCCA and DCL use deep neural networks (DNN) for feature extraction. DNN employs a sequence of intermediate layers to extract high-level features. Each layer is composed of a number of complex operations, e.g., nonlinear activation, kernel convolution, batch normalization. DNN based models have found numerous successful applications in both computer vision and medical imaging fields, as a result of their superior ability to extract high-level features. However, the large number of layers and the complex/nonlinear operations in each layer bring about a difficulty in network explanation and feature identification. As a result, users may cast doubt on the reliability of the deep networks: whether deep networks make decisions based on the object of interest, or based on irrelevant/background information.
\subsubsection{Class Activation Mapping (CAM)}
To make DNN explainable, Class Activation Mapping (CAM) method \cite{zhou2016learning} was proposed. CAM generates an activation map for each sample/image indicating pixel-wise contributions to the decision of interest, e.g., class label. Moreover, as its name tells, CAM's activation maps are class-specific, providing more discriminative information for further class-specific analysis. This dramatically helps build trust in deep networks: for correctly classified images/samples, CAM explains how the classification decision is made by highlighting the object of interest; for incorrectly classified images/samples, CAM illustrates why incorrect decisions are made by highlighting the misleading regions.

CAM's activation maps are obtained by computing an optimal combination of intermediate feature maps. As feature maps only exist in convolutional layers, CAM can be applied only to convolutional neural networks (CNN). A weight coefficient is needed for each feature map to evaluate its importance to the decision of interest. However, for most CNN based models, this weight is not provided. To solve this problem, a re-training procedure is introduced, in which the feature maps are used directly by a newly introduced layer to re-conduct classification. The weights can then be calculated using the parameters in the introduced layer accordingly. The detailed CAM method is described as follows.

For a pre-trained CNN-based model, assume that a target feature map layer consists of $K$ channels/feature-maps $F^{k} \in \mathbb{R}^{h \times w} (k = 1, 2, \cdots, K)$, where $h, w$ represent the height and width of each feature map, respectively. CAM discards all the subsequent layers and then introduces a new layer (with softmax activation) to re-conduct classification using these feature maps $F^{k}$. A prediction score $S^{c}$  will then be calculated by the newly introduced layer for each class $c \;(c = 1, 2, \cdots, C)$, as formulated in Eq. \ref{eq:cam_predict}.
\begin{align}
S^{c} = \sum_{k=1}^{K} w^{c}_{k} \; \mbox{global\_avg\_pooling}\hspace{-0.25em}\left(F^{k}\right) \label{eq:cam_predict} 
\end{align}
where $w^{c}_{k}$ represents the weight coefficient of feature map $F^{k}$ for class $c$.

After that, class-specific activation maps can be generated by first combining the feature maps using the trained weights $w^{c}_{k}$ and then conducting upsampling to project it onto input images, as in Eq. \ref{eq:cam_map}.
\begin{align}
\mbox{map}_{cam} = \mbox{upsampling}\hspace{-0.35em}\left( \sum_{k=1}^{K} w^{c}_{k}F^{k}  \right) \label{eq:cam_map}
\end{align}

The re-training procedure, however, is time consuming, which limits CAM's application. Moreover, classification accuracy will sacrifice due to the modification of the model's architecture, and consequently the accuracy of activation maps will decrease. 

\subsubsection{Gradient-weighted CAM (Grad-CAM)}
To address the limitations of the CAM method, Gradient-weighted CAM (Grad-CAM), was proposed \cite{selvaraju2017grad} to compute activation maps without modifying the model's architecture. Similar to CAM, Grad-CAM also needs a set of weight coefficients so as to combine feature maps. This can be achieved by first calculating the gradients of decision of interest w.r.t each feature maps and then performing global average pooling on the gradients to get scalar weights. In this way, Grad-CAM avoids adding extra layers and consequently both model-retraining and performance-decrease problems can be solved. The formulations of how Grad-CAM calculates weights $g^{c}_{k}$ and activation map $\mbox{map}_{gradcam}$ are as follows.
\begin{align}
g^{c}_{k} = \mbox{global\_avg\_pooling}\hspace{-0.3em} \left( \frac{\partial y^c}{\partial F^k} \right) \label{eq:weight_gcam}
\end{align}
where $y^c$ represents the prediction score for class $c$, and
\begin{align}
\mbox{map}_{gradcam} = \mbox{upsampling}\hspace{-0.35em}\left( \sum_{k=1}^{K} g^{c}_{k}F^{k}  \right) \label{eq:gcam_map}
\end{align}

\subsubsection{Guided Grad-CAM: high resolution class-specific activation maps}
Both CAM's and Grad-CAM's activation maps are coarse due to the upsampling procedure, as feature maps normally are of smaller size compared to input images. This brings about difficulties in identifying small but important object-features. Fine-grained visualization methods, e.g., guided backpropagation (BP) \cite{springenberg2014striving} and deconvolution \cite{zeiler2014visualizing}, can generate high resolution activation maps. These methods use backward projections which operate on layer-to-layer gradients. Upsampling procedure is not involved in these back projection methods, and therefore high resolution activation maps can be obtained. Nevertheless, the activation maps are not class-specific, bringing about obstacles in interpreting the activation maps, especially for multi-class (more than 2) scenarios. To obtain both high resolution and class-specific activation maps, guided Grad-CAM was proposed in the work \cite{selvaraju2017grad} which incorporated guided BP into Grad-CAM. Guided Grad-CAM computes activation maps by performing a Hadamard product between the Grad-CAM map and the Guided BP map, as formulated in Eq. \ref{eq:map_guided_gcam}.
\begin{align}
\mbox{map}_{guided\_gradcam} = \mbox{map}_{guidedBP} \odot \mbox{map}_{gradcam} \label{eq:map_guided_gcam}
\end{align}
where $\mbox{map}_{guidedBP}$ represents the map computed using guided BP algorithm \cite{springenberg2014striving}, and $\odot$ represents the Hadamard product operation. For example, given two arbitrary matrices $A, B \in \mathbb{R}^{m \times n}$, their Hadamard product is defined as $(A \odot B)_{ij} := A_{ij}B_{ij}$.

\subsection{Grad-CAM guided convolutional collaborative learning (gCAM-CCL)}
For the purpose of interpretable multimodal fusion, we develop a new model, Grad-CAM guided convolutional collaborative learning (gCAM-CCL), which incorporates both guided BP and Grad-CAM methods into the DCL model. As shown in Fig. \ref{fig:gcam-ccl_net}, gCAM-CCL first integrates two modality data using the collaborative networks, and then computes class-specific activation maps using Guided BP and Grad-CAM. In this way, gCAM-CCL can perform both automated classification/diagnosis and automated biomarker-identification as well as result interpretation simultaneously.

To be more specific, gCAM-CCL uses a 1D ConvNet to learn features from SNP data and uses a 2D ConvNet to learn features from brain imaging data. The output of two ConvNets are flattened and then fused in the collaborative layer with the loss function in Eq. \ref{eq:lossfun}, which considers both cross-data associations and their fittings to phenotype/label $y$. After that, two intermediate layers will be selected, from which the feature maps will be combined using the gradient-based weights (Eqs. \ref{eq:weight_gcam}-\ref{eq:gcam_map}) and class-specific Grad-CAM activation maps will be generated accordingly. Meanwhile, fine-grained activation maps are computed by projecting the gradients back from the collaborative layer to the input layer using Guided BP. The obtained activation maps indicate pixel-wise contributions to the decision of interest, e.g., prediction, and significant biomarkers, e.g., brain FCs and genes, can be identified accordingly.

Compared to the DCL model \cite{hu2019deep}, gCAM-CCL employs both new architecture and new loss function so as to incorporate Grad-CAM. As computing activation maps needs a layer of feature-maps, gCAM-CCL replaces a multilayer perceptron (MLP) network with two ConvNets so that multi-channel feature maps can be obtained. This also benefits model-training as ConvNet dramatically reduces the the number of parameters by enforcing shared kernel weights.

Moreover, to compute class-specific activation maps, gradients $\frac{\partial y^c}{\partial F^k}$ w.r.t. each class $c \;(c = 1, 2, \cdots, C)$ are needed, as illustrated in Eq. \ref{eq:weight_gcam}.. However, DCL uses external classifiers, e.g., support vector machine (SVM), and therefore no class information is provided in DCL's gradients. To solve this problem, gCAM-CCL replaces external classifiers with an embedded softmax classifier so that class-specific gradients $g_c^k$ can be obtained.

Furthermore, ideal class-specific activation maps should highlight only the features relevant to the corresponding class, e.g., 'dog' class. However, features related to other classes, e.g.,  fish-related features,  may have strong but negative contributions to predicting 'dog' class, resulting in noise features in the activation maps. To remove the noise features, we apply a ReLU function to the gradients, as shown in Eq. \ref{eq:weight_gcam_relu}. The ReLU function ensures positive effects so that pixels with negative contributions can be filtered out.
\begin{align}
g^{c}_{k} = \mbox{global\_avg\_pooling}\hspace{-0.3em} \left( ReLU(\frac{\partial y^c}{\partial F^k}) \right) \label{eq:weight_gcam_relu}
\end{align}
where $y^c$ represents the prediction score for class $c$.

In addition, as pointed out in Wang's work \cite{wang2015deep}, both DCCA\cite{deepcca} and DCL \cite{hu2019deep} include the parameter of sample size into their loss functions, resulting in a problem in batch size tuning. In other words, their loss functions are dependent on batch size due to a population-level correlation term $U_1'Z_1'Z_2U_2$. As a result, a large batch size is required \cite{wang2015deep}, leading to a challenge for batch size tuning and network training. In this work, we propose a new loss function which resolves the batch-size dependence, as formulated in Eq. \ref{eq:lossfun}. As shown in Eq. \ref{eq:lossfun}, the population-level correlation term is replaced with a summation of sample-level loss. Moreover, the correlation term is replaced with a regression loss, i.e., cross-entropy loss, as it has been shown that the optimization of correlation term is equivalent to the optimization of regression loss \cite{wang2015deep}.
\begin{align}
    Loss = & -\sum_{i=1}^{2} \left( (1-y)log(h_1^{(i)}) + ylog(1-h_2^{(i)}) \right) \label{eq:lossfun} \\
    & -\sum_{i=1}^{2} \left( h_i^{(1)}log(h_i^{(2)}) + h_i^{(2)}log(h_i^{(1)})\right) \notag
\end{align}
where $h_{i}^{(1)}, h_{i}^{(2)}$ are the outputs of two ConvNets, as illustrated in Fig. \ref{fig:gcam-ccl_net}.

\noindent This batch-independent loss function is easier to extend to multi-class multi-view scenarios and the extended loss function is formulated as follows.
\begin{align}
    Loss = & -\frac{1}{m} \sum_{i=1}^m \sum_{c=1}^C y_c log(h_c^{(i)}) \label{eq:lossfun_multi}\\
 & -\frac{1}{m(m-1)} \sum_{i,j (i\neq j)}^{m} \sum_{c=1}^C h_c^{(i)} log( h_c^{(j)}) \notag
\end{align}
where $m$ represents the number of views, and $C$ represents the number of classes.

\section{Application to brain imaging genetic study}
We apply the gCAM-CCL model to an imaging genetic study, in which brain FC data is integrated with single nucleotide polymorphism (SNPs) data to classify low/high cognitive groups. Multiple brain regions of interests (ROIs) function as a group when performing a specific task, e.g., reading. Brain FC depicts the functional associations between different brain ROIs \cite{connectivity_vince}. On the other hand, genetic factors may also have influences on brain functions, as brain dysfunctionality is genetically inheritable. Imaging-genetic integration enables exploring brain function from a more comprehensive view, which may further contribute to the study of normal and pathological brain mechanisms. The proposed gCAM-CCL model, which can perform automated diagnosis and feature interpretation, can be used to extract and analyze the complex interactions both within and between brain FC data and genetic data.

\subsection{Brain imaging data}
Several brain fMRI modalities from the Philadelphia Neurodevelopmental Cohort (PNC) \cite{satterthwaite2014neuroimaging} were used in the experiments. PNC cohort is a large-scale collaborative study between the Brain Behavior Laboratory at the University of Pennsylvania and the Children’s Hospital of Philadelphia. It has a collection of multiple neuroimaging data, e.g., fMRI, and genomic data, e.g., SNPs, from adolescents aged from 8 to 21 years. Three types of fMRI data are available in PNC cohort: resting-state fMRI, emotion task fMRI, and nback task fMRI (nback-fMRI). As our work was focused on analyzing cognitive ability, only nback-fMRI, which was related to working memory and lexical processing, was used in the experiments. The duration of nback-fMRI scan was 11.6 minutes (231 TR), during which subjects were asked to conduct standard nback tasks.

SPM12\footnote{http://www.fil.ion.ucl.ac.uk/spm/software/spm12/} was used to conduct motion correction, spatial normalization, and spatial smoothing. Movement artefact (head motion effect) was removed via a regression procedure using a rigid body (6 parameters: 3 translation and 3 rotation parameters) \cite{friston1995characterizing}, and the functional time series were band-pass filtered using a 0.01Hz to 0.1Hz frequency range as significant signals mainly focus on low frequency. For quality control, we excluded high motion subjects with translation $>$ 2mm or with SFNR $<$ 275 (Signal-to-fluctuation-noise ratio) following the work in \cite{rashid2014dynamic}. 264 regions of interest (ROIs) (containing 21,384 voxels) were extracted based on the Power coordinates \cite{power} with a sphere radius parameter of 5mm. For each subject, a $264\times 264$ image was then obtained based on the $264\times 264$ ROI-ROI connections, which was used next as image inputs for the gCAM-CCL model.
\subsection{SNP data}
The genomic data were collected from 3 platforms, including the Illumina HumanHap 610 array, the Illumina HumanHap 500 array, and the Illumina Human Omni Express array. The three platforms generated 620k, 561k, 731k SNPs, respectively \cite{satterthwaite2014neuroimaging}. A common set of SNPs (313k) were extracted, and then PLINK \cite{purcell2007plink} was used to perform standard quality controls, including the Hardy-Weinberg equilibrium test for genotyping errors with p-value $<$ 1e$-$5, extraction of common SNPs (MAF $>$ 5\%), and linkage disequilibrium (LD) pruning with a threshold of 0.9. After that, SNPs with missing call rates $>$ 10\% and samples with missing SNPs $>$ 5\% were removed. The remaining missing values were imputed by Minimac 3 \cite{das2016next} using the reference genome from 1000 Genome Project. In addition, only the SNPs within gene bodies were kept for further analysis, resulting in 98,804 SNPs in 14,131 genes.

As the study aimed to investigate the brain, we further narrowed down the scope to brain-expression-related SNPs. This was achieved using the expression quantitative trait loci (eQTL) data from Genotype-Tissue Expression (GTEx)\footnote{https://gtexportal.org/} database \cite{lonsdale2013genotype}, a large scale consortium studying tissue-specific gene regulations and expressions. The GTEx data were collected from 53 different tissue sites from around 1000 subjects. Among the 53 tissue sites, 13 tissues were brain-related and they were listed in Table \ref{tab:brain_tissue}. A set of 108 SNP loci which showed significant tissue regulation level (eQTL $<5\times10e$-8) in all 13 brain relevant tissues were selected. In addition, SNPs in the top 100 brain-expressed genes were also selected based on the GTEx database. These procedures resulted in 750 SNP loci, which were used next as the genetic input for the gCAM-CCL model.
\begin{table}[pt]
\caption{13 Brain relevant tissues from GTEx database}
\label{tab:brain_tissue}
\begin{center}
\begin{tabular}{l l} 
\hline \hline
Brain amygdata& Brain nucleus accumbens\\
Brain caudate& Brain cerebellar hemisphere\\
Brain cerebellum& Brain frontal cortex\\
Brain cortex& Brain substantisa nigra\\
Brain putamen& Brain anterior cingulate cortex\\
Brain spinal cord& Brain hypothalamus\\
Brain hippocampus\\
\hline
\end{tabular}
\end{center}
\end{table}

\subsection{Integrating brain imaging and genetic data: classification}
The gCAM-CCL was then applied to integrate brain imaging data with SNPs data to classify subjects with low/high cognitive abilities. The wide range achievement test (WRAT)\cite{wrat} score, a measure of comprehensive cognitive ability, including reading, comprehension, math skills, etc., was used to evaluate the cognitive ability of each subject. The 854 subjects were divided into three classes: high cognitive/WRAT group (top 20\% WRAT score), low cognitive/WRAT group (bottom 20\% WRAT score), and middle group (the rest), following the procedures in work \cite{hu2019deep}.

The gCAM-CCL model adopted a 1D convolutional nieural network (CNN) to learn the interactions between alleles at different single-nucleotide polymorphism (SNP) loci. ConvNet has been widely used on sequencing and gene expression data \cite{singh2016deepchrome,zhou2018deep} to learn local genetic structures. According to these studies, 1D kernels with relatively larger size are preferred. As a result, a $31\times1$ kernel and a $15\times1$ kernel were used. The detailed architecture of gCAM-CCL is listed in Table \ref{tab:architect}. The partition of the data is as follows: training set (70\%), validation set (15\%), and test set (15\%). The proposed gCAM-CCL model was trained on training set; hyper-parameters were selected based on the loss on the validation set; and the classification performance was reported based on the test set.

Hyper-parameters, including momentum, activation function, learning rate, decay rate, batch size, maximum epochs, were selected using the validation set and their values were listed in Table \ref{tab:hyperpara}. Mini-batch SGD was used to solve the optimization problem. Over-fitting problem occurred due to small sample size. To solve overfitting, dropout was used and the dropout probability of the middle layers was set to be 0.2. Moreover, early stopping was used during network training to further address overfitting. In addition, batch normalization was implemented after each layer to relieve the gradient vanishing/exploding problem resulting from ReLU activation. Computational experiments were conducted on a Desktop with an Intel(R) Core(TM) i7-8700K CPU (@ 3.70GHz), a 16G RAM, and a NVIDIA GeForce GTX 1080 Ti GPU (11G).

\begin{table*}[ht]
\caption{Hyper-parameter setting}
\label{tab:hyperpara}
\begin{center}       
\begin{tabular}{l l l l l l l l} 
\hline \hline
\textbf{Methods} & \textbf{Epochs} & \textbf{batch size} & \textbf{Activation} & \textbf{Learning rate} & \textbf{Decay rate} & \textbf{dropout} & \textbf{Momentum} \\
\hline 
gCAM-CCL & 500 & 4 & ReLU, Sigmoid & 0.00001 & Half per 200 epochs & 0.2 (middle layers) & 0.9\\
\hline
\end{tabular}
\end{center}
\end{table*}

\begin{table}
\caption{The comparison of classification performances (Low/High WRAT classification).}
\label{tab:wrat_acc}
\begin{center}
\begin{tabular}{l l l l l l} 
\hline \hline
\textbf{Classifier}&\textbf{ACC}&\textbf{SEN}&\textbf{SPF}&\textbf{F1}\\
\hline
gCAM-CCL & 0.7501 & 0.7762 & 0.7157 & 0.7610\\
CCL+SVM & 0.7387 & 0.7637 & 0.7083 & 0.7504\\
CCL+RF &  0.7419 & 0.7666 & 0.7014 & 0.7523\\
MLP+SVM & 0.7231 & 0.7555 & 0.6915 & 0.7215\\
SVM & 0.7082 & 0.7562 & 0.6785 & 0.7093\\
DT & 0.6626 & 0.6778 & 0.6430 & 0.6605\\
RF & 0.7119 & 0.7559 & 0.6714 & 0.7138\\
Logist & 0.6745 & 0.7386 & 0.6285 & 0.6900\\
\hline
\end{tabular}
\end{center}
\end{table}

For the purpose of comparison, several classical classifiers, e.g., SVM, random forest (RF), decision tree,  were implemented for classifying low/high WRAT groups. In addition, several deep network based classifiers were implemented, including CCL with external classifiers (SVM/RF), multilayer perceptron (MLP). The result of classifying high/low cognitive groups was shown in Table \ref{tab:wrat_acc}. From Table \ref{tab:wrat_acc}, gCAM-CCL outperforms both conventional classifiers, e.g., SVM, and regular deep network fusion method, in which two data were concatenated as the input. This is consistent with the result in the work \cite{hu2019deep}, which also showed that the collaborative network can improve classification performance for multimodal data.  Moreover, gCAM-CCL with intrinsic softmax classifiers achieved better classification performance compared with 'CCL+SVM' and 'CCL+RF'. This may be due to the incorporation of cross-entropy loss, i.e., Eq. \ref{eq:lossfun}, which helps the network more efficiently learn loss-gradient during back-propagation process at each iteration.

\begin{table*}[htbp]
\caption{The Architecture of gCAM-CCL}
\label{tab:architect}
\begin{center}
\begin{tabular}{l l l l l l l l} 
\hline
\multicolumn{3}{c}{fMRI ConvNet} & & \multicolumn{3}{c}{SNP ConvNet} & \\
\hline
Layer Name      & Input Shape   & Operations    & Connects to \qquad \qquad \qquad  & Layer Name    & Input Shape   & Operations    & Connects to \\
\hline
f\_conv1        & (b, 1, 264, 264)      & K, P, MP = 7, 3, 2    & f\_conv2      & s\_conv1        & (b, 1, 750)   & K, MP = 31, 6 & s\_conv2 \\
f\_conv2        & (b, 16, 132, 132)     & K, P, MP = 5, 2, 4    & f\_conv3      & s\_conv2        & (b, 16, 120)  & K, MP = 31, 6 & s\_conv3 \\
f\_conv3        & (b, 32, 33, 33)       & K, P, MP = 3, 1, 3    & f\_conv4      & s\_conv3        & (b, 32, 15)   & K = 15        & s\_flatten \\
f\_conv4        & (b, 32, 11, 11)       & K = 11        & f\_flatten    & s\_flatten      & (b, 64, 1)    &-      & collab\_layer \\
f\_flatten      & (b, 64, 1)    &-      & collab\_layer &-&-&-&-\\
\hline
collab\_layer   & (b, 4) &&&&&& \\
\hline
\multicolumn{4}{l}{Notations: b (batch size), K (kernel size), P (padding), MP (maxpooling).}
\end{tabular}
\end{center}
\end{table*}

\begin{table}[htbp]
\caption{Identified SNP loci (Low WRAT group)}
\label{tab:gene_class0}
\begin{center}
\begin{tabular}{l l l l} 
\hline \hline
SNP rs \# & Gene & SNP rs \# & Gene \\
\hline
rs1642763       & ATP1B2        & rs997349      & MTURN \\
rs9508      & ATPIF1    & rs17547430    & MTURN \\
rs2242415       & BASP1 & rs7780166     & MTURN \\
rs11133892      & BASP1 & rs10488088    & MTURN \\
rs10113     & CALM3     & rs3750089     & MTURN \\
rs11136000      & CLU   & rs2275007     & OSGEP \\
rs4963126       & DEAF1 & rs4849179     & PAX8 \\
rs11755449      & EEF1A1        & rs11539202    & PDHX \\
rs2073465       & EEF1A1        & rs1045288     & PSMD13 \\
rs1809148       & EEF1D & rs7563960     & RNASEH1 \\
rs4984683       & FBXL16        & rs145290      & RP1 \\
rs7026635       & FBXW2 & rs446227      & RP1 \\
rs734138        & FLYWCH1       & rs414352      & RP1 \\
rs2289681       & GFAP  & rs6507920     & RPL17 \\
rs7258864       & GNG7  & rs12484030    & RPL3 \\
rs4807291       & GNG7  & rs10902222    & RPLP2 \\
rs887030        & GNG7  & rs8079544     & TP53 \\
rs7254861       & GNG7  & rs6726169     & TTL \\
rs12985186      & GNG7  & rs415430      & WNT3 \\
rs2070937       & HP    & rs8078073     & YWHAE \\
rs622082        & IGHMBP2       & rs12452627    & YWHAE \\
rs12460     & LINS      & rs324126      & ZNF880 \\
rs10044354      & LNPEP & & \\
\hline
\end{tabular}
\end{center}
\end{table}

\begin{table}[htbp]
\caption{Identified SNP loci (High WRAT group)}
\label{tab:gene_class1}
\begin{center}
\begin{tabular}{l l l l} 
\hline \hline
SNP rs \# & Gene & SNP rs \# & Gene \\
\hline
rs3787620       & APP   & rs1056680         & MB \\
rs373521        & APP   & rs9257936         & MOG \\
rs2829973       & APP   & rs7660424         & MRFAP1 \\
rs1783016       & APP   & rs3802577         & PHYH \\
rs440666        & APP   & rs1414396         & PHYH \\
rs2753267       & ATP1A2        & rs1414395     & PHYH \\
rs10494336      & ATP1A2        & rs1037680     & PKM \\
rs1642763       & ATP1B2        & rs2329884     & PPM1F \\
rs10113     & CALM3     & rs1045288     & PSMD13 \\
rs2053053       & CAMK2A        & rs2271882     & RAB3A \\
rs4958456       & CAMK2A        & rs12294045    & SLC1A2 \\
rs4958445       & CAMK2A        & rs3794089     & SLC1A2 \\
rs3756577       & CAMK2A        & rs7102331     & SLC1A2 \\
rs874083        & CAMK2A        & rs3798174     & SLC22A1 \\
rs3011928       & CAMTA1        & rs9457843     & SLC22A1 \\
rs890736        & CPLX2 & rs1443844     & SLC22A1 \\
rs17065524      & CPLX2 & rs6077693     & SNAP25 \\
rs12325282      & FAHD1 & rs363043      & SNAP25 \\
rs104664        & FAM118A       & rs362569      & SNAP25 \\
rs6874      & FAM69B    & rs10514299    & TMEM161B-AS1 \\
rs7026635       & FBXW2 & rs4717678     & TYW1B \\
rs12735664      & GLUL  & rs8078073     & YWHAE \\
rs7155973       & HSP90AA1      & rs10521111    & YWHAE \\
rs2251110       & LOC101928134  & rs4790082     & YWHAE \\
rs2900856       & LOC441242     & rs10401135    & ZNF559 \\
rs8136867       & MAPK1 & & \\
\hline
\end{tabular}
\end{center}
\end{table}

\begin{table*}[htbp]
\caption{Gene enrichment analysis of the identified genes (Low WRAT group). Q-values represent multiple testing corrected p-value.}
\label{tab:pathway_class0}
\begin{center}
\begin{tabular}{l l l l l l} 
\hline \hline
Pathway Name    & Pathway Source        & Set size      & Contained     & p-value & q-value \\
\hline
Eukaryotic Translation Elongation       & Reactome      & 106   & 5     & 1.18E-06        & 1.65E-05 \\
Peptide chain elongation        & Reactome      & 101   & 4     & 3.12E-05      & 2.18E-04 \\
Calcium Regulation in the Cardiac Cell  & Wikipathways  & 149   & 4     & 1.48E-04        & 6.89E-04 \\
Translation     & Reactome      & 310   & 5     & 2.09E-04      & 7.33E-04 \\
Metabolism of proteins  & Reactome      & 2008  & 11    & 3.96E-04      & 1.11E-03 \\
Midbrain development    & Gene Ontology & 94    & 4     & 1.22E-05      & 1.53E-03 \\
Site of polarized growth        & Gene Ontology & 167   & 4     & 1.25E-04      & 2.19E-03 \\
Growth cone     & Gene Ontology & 165   & 4     & 1.20E-04      & 4.07E-03 \\
Cellular catabolic process      & Gene Ontology & 2260  & 12    & 7.22E-05      & 4.55E-03 \\
Pathways in cancer - Homo sapiens (human)       & KEGG  & 526   & 5     & 2.39E-03        & 5.58E-03 \\
Metabolism of amino acids and derivatives       & Reactome      & 342   & 4       & 3.20E-03      & 6.40E-03 \\
\hline
\end{tabular}
\end{center}
\end{table*}

\begin{table*}[htbp]
\caption{Gene enrichment analysis of the identified genes (High WRAT group). Q-values represent multiple testing corrected p-value.}
\label{tab:pathway_class1}
\begin{center}
\begin{tabular}{l l l l l l} 
\hline \hline
Pathway Name    & Pathway Source        & Set size      & Contained     & p-value & q-value \\
\hline
Regulation of neurotransmitter levels   & Gene Ontology & 335   & 9     & 6.77E-10        & 1.04E-07 \\
Transmission across Chemical Synapses   & Reactome      & 224   & 7     & 7.75E-08        & 2.40E-06 \\
Synaptic signaling      & Gene Ontology & 711   & 10    & 3.17E-08      & 2.43E-06 \\
Insulin secretion - Homo sapiens (human)        & KEGG  & 85    & 5     & 3.26E-07        & 5.06E-06 \\
Organelle localization by membrane tethering    & Gene Ontology & 170   & 6       & 1.34E-07      & 6.84E-06 \\
Regulation of synaptic plasticity       & Gene Ontology & 179   & 6     & 1.89E-07        & 7.21E-06 \\
Exocytosis      & Gene Ontology & 909   & 10    & 3.06E-07      & 9.36E-06 \\
Membrane docking        & Gene Ontology & 179   & 6     & 1.82E-07      & 1.28E-05 \\
Vesicle docking involved in exocytosis  & Gene Ontology & 45    & 4     & 5.11E-07        & 1.30E-05 \\
Plasma membrane bounded cell projection part    & Gene Ontology & 1452  & 12      & 3.05E-07      & 1.34E-05 \\
Cell projection part    & Gene Ontology & 1452  & 12    & 3.05E-07      & 1.37E-05 \\
Neurotransmitter release cycle  & Reactome      & 51    & 4     & 1.76E-06      & 1.37E-05 \\
Synaptic Vesicle Pathway        & Wikipathways  & 51    & 4     & 1.76E-06      & 1.37E-05 \\
Neuronal System & Reactome      & 368   & 7     & 2.25E-06      & 1.39E-05 \\
Secretion by cell       & Gene Ontology & 1493  & 12    & 4.13E-07      & 1.45E-05 \\
Adrenergic signaling in cardiomyocytes - Homo sapiens (human)   & KEGG  & 144     & 5     & 4.47E-06      & 2.31E-05 \\
Gastric acid secretion - Homo sapiens (human)   & KEGG  & 75    & 4     & 8.34E-06        & 3.70E-05 \\
Neuron part     & Gene Ontology & 1713  & 12    & 1.83E-06      & 4.09E-05 \\
Plasma membrane bounded cell projection & Gene Ontology & 2098  & 13    & 2.22E-06        & 4.87E-05 \\
\hline
\end{tabular}
\end{center}
\end{table*}

\begin{figure}[pt]
\begin{center}
\begin{tabular}{c}
\includegraphics[width=0.46\textwidth]{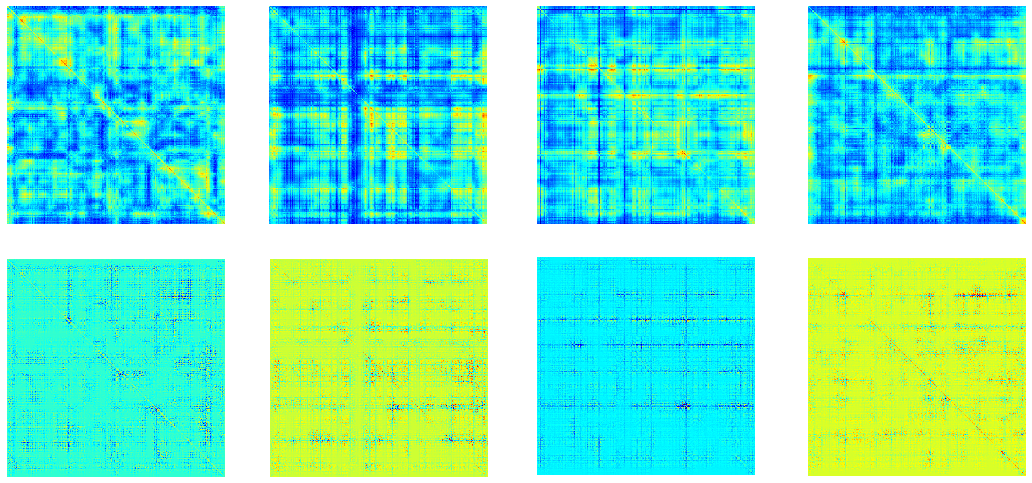}
\end{tabular}
\end{center}
\caption 
{The brain FC activation maps for Low WRAT group: Grad-CAM (top 4 subfigures) and Gradient-Guided Grad-CAM (bottom 4 subfigures).} \label{fig:gradcam_class0}
\end{figure}

\begin{figure}[pt]
\begin{center}
\begin{tabular}{c}
\includegraphics[width=0.46\textwidth]{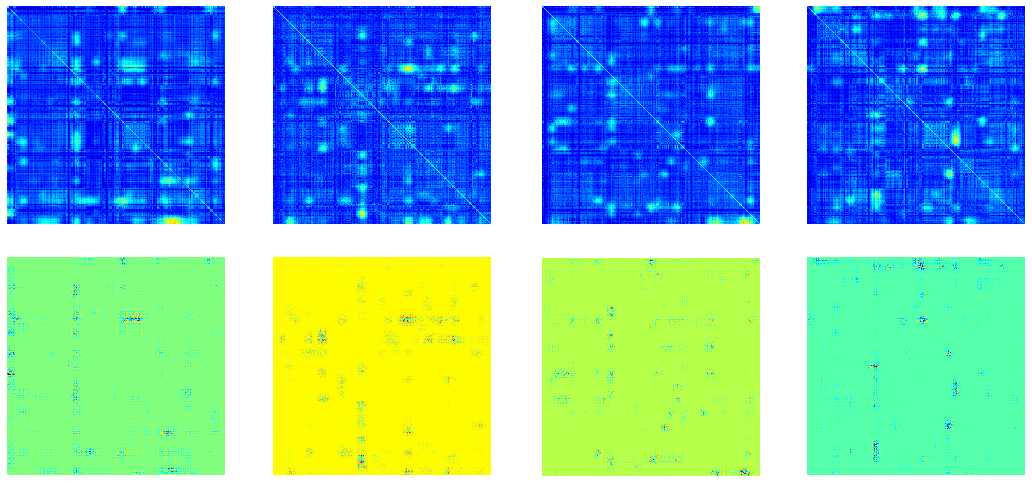}
\end{tabular}
\end{center}
\caption 
{The brain FC activation maps for High WRAT group: Grad-CAM (top 4 subfigures) and Gradient-Guided Grad-CAM (bottom 4 subfigures).} \label{fig:gradcam_class1}
\end{figure}

\begin{figure}
\begin{center}
\begin{tabular}{c}
\includegraphics[width=0.46\textwidth]{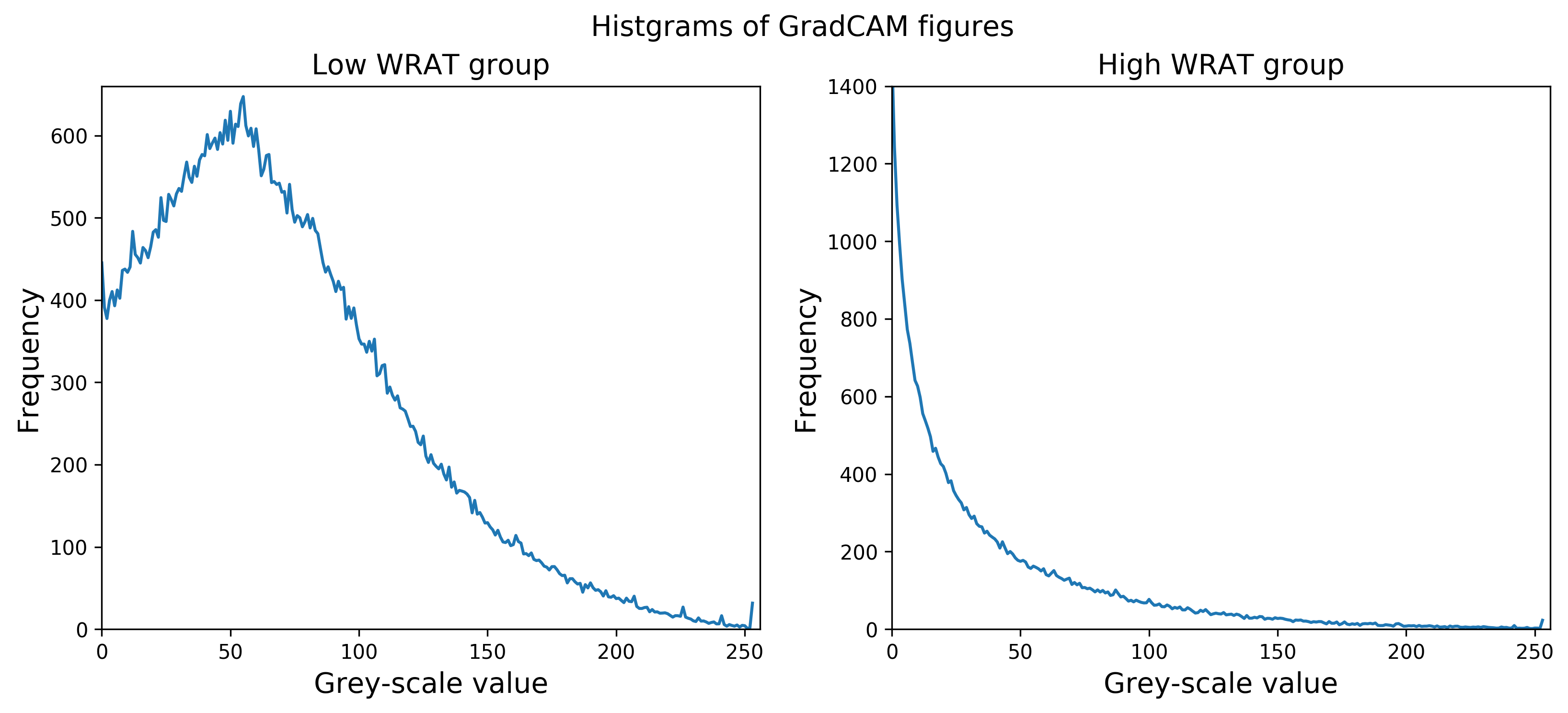}
\end{tabular}
\end{center}
\caption{The histogram of the Grad-CAM activation maps of brain FCs (see Figs. \ref{fig:gradcam_class0}-\ref{fig:gradcam_class1}).} \label{fig:hist}
\end{figure}

\begin{figure*}
\begin{center}
\begin{tabular}{c}
\includegraphics[width=0.9\textwidth]{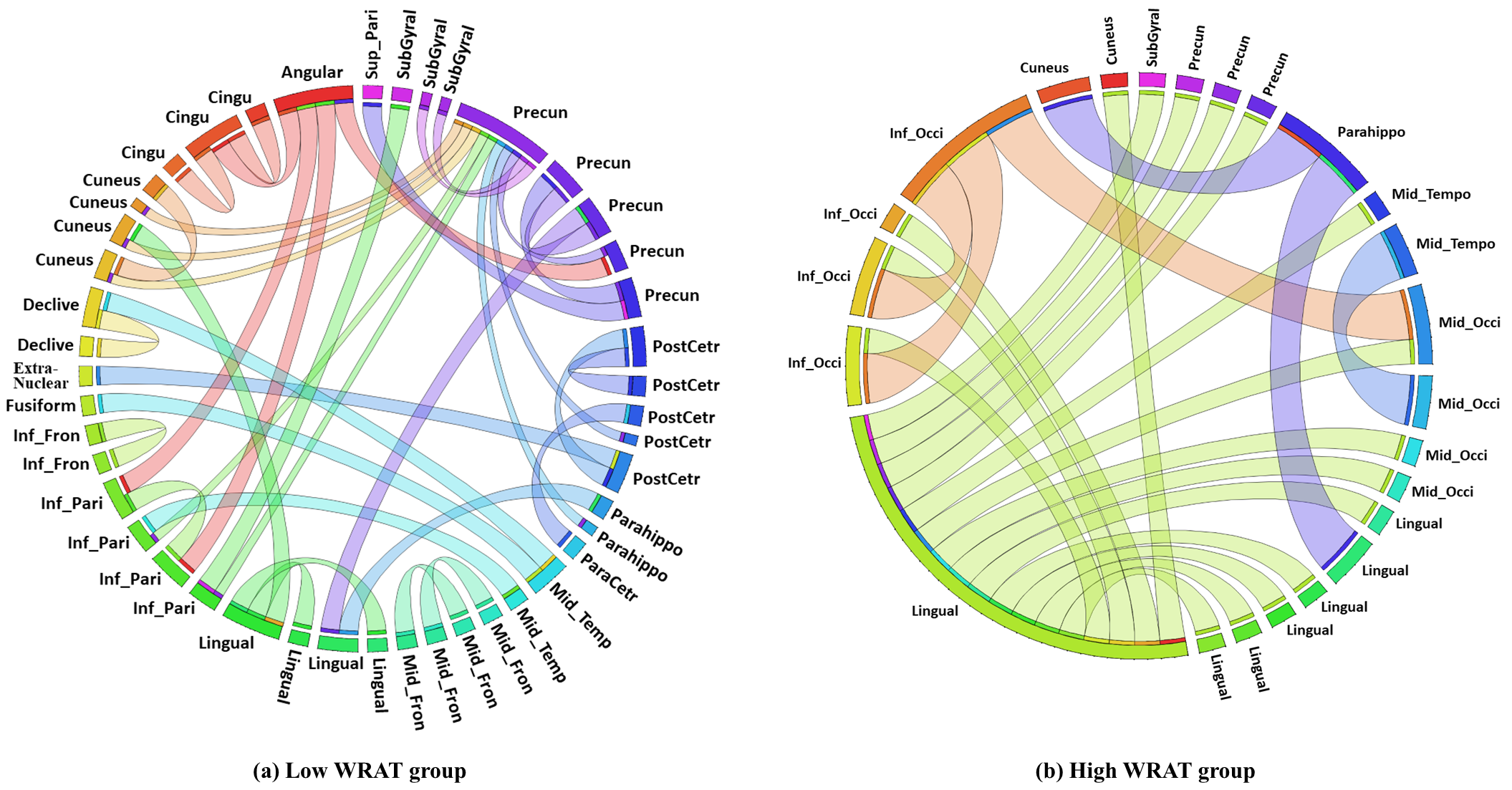}
\end{tabular}
\end{center}
\caption{The identified class-discriminative brain FCs by gCAM-CCL. The full names of ROIs can be found in Table \ref{tab:roinames_abbre}. Each circle arc represents a ROI (based on Power parcellation \cite{power}). The length of a circle arc indicates the number of ROI-ROI connections on this ROI.}\label{fig:fc_circle}
\end{figure*}

\begin{figure}
\begin{center}
\begin{tabular}{c}
\includegraphics[width=0.46\textwidth]{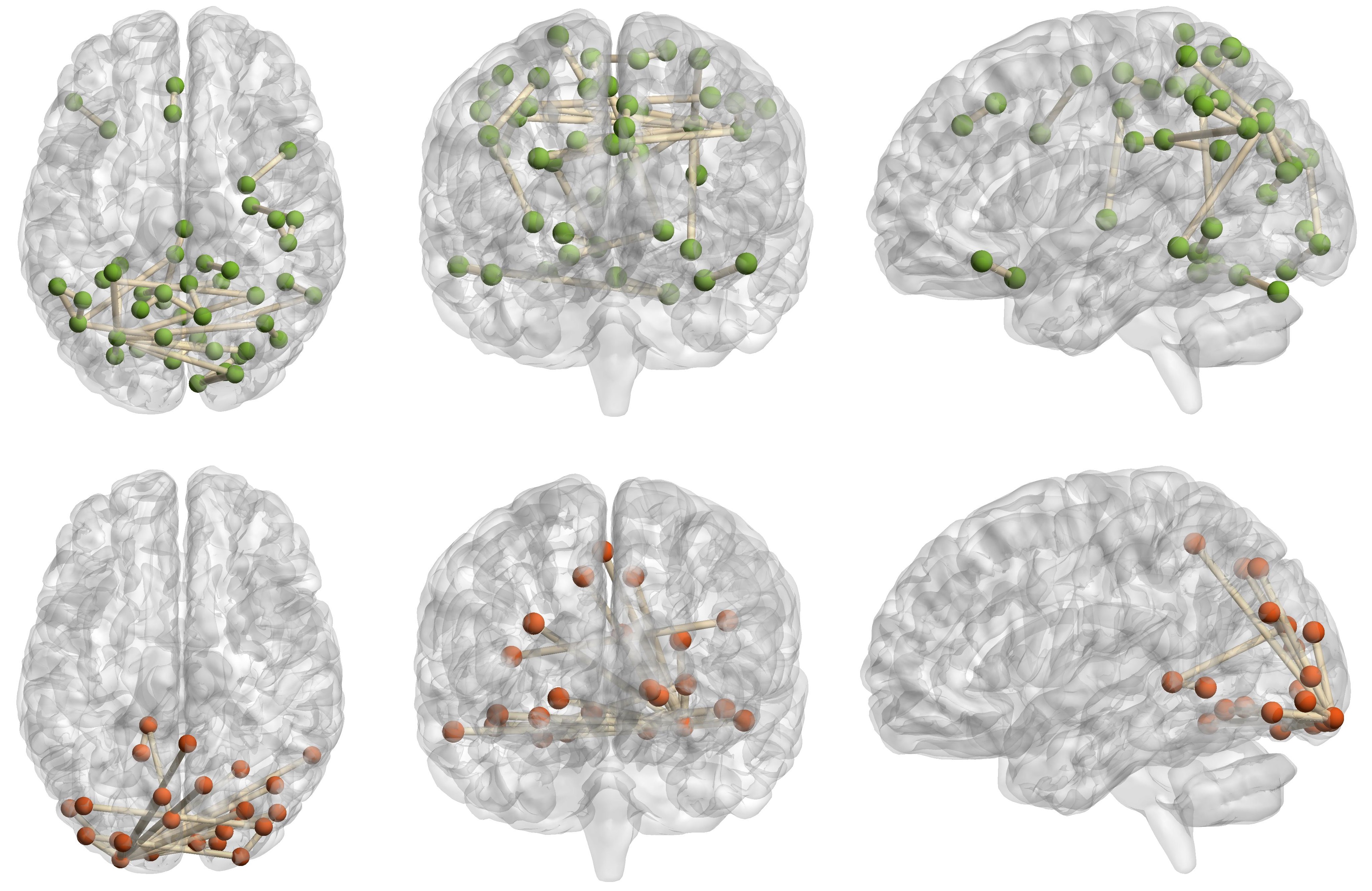}
\end{tabular}
\end{center}
\caption{The identified brain functional connectivity. The top 3 subfigures: Low WRAT group (axial view, coronal view, sagittal view, respectively); the bottom 3 subfigures: High WRAT group (axial view, coronal view, sagittal view, respectively).} \label{fig:fc_graph}
\end{figure}

\begin{table}
\caption{Abbreviations of the ROIs}
\label{tab:roinames_abbre}
\begin{center}
\begin{tabular}{l l} 
\hline \hline
Inferior Parietal Lobule (Inf\_Pari) & Angular Gyrus (Angular)\\
Inferior Occipital Gyrus (Inf\_Occi) & Fusiform Gyrus (fusiform)\\
Inferior Frontal Gyrus (Inf\_Fron) & Cingulate Gyrus (Cingu)\\
Middle Occipital Gyrus (Mid\_Occi) & Sub-Gyral (SubGyral)\\
Middle Frontal Gyrus (Mid\_Fron) & Paracentral Lobule (ParaCetr)\\
Parahippocampa Gyrus (Parahippo) & Postcentral Gyrus (PostCetr)\\
Middle Temporal Gyrus (Mid\_Temp) & Precuneus (Precun)\\
Superior Parietal Lobule (Sup\_Pari)\\
\hline
\end{tabular}
\end{center}
\end{table}

\subsection{Integrating brain imaging and genetic data: result interpretation}
The class-specific activation maps for low WRAT group and high WRAT group were plotted in Figs. \ref{fig:gradcam_class0}-\ref{fig:gradcam_class1}, respectively. From Fig. \ref{fig:gradcam_class0}, the low WRAT group shows a relatively larger number of activated FCs, which contributed to making the 'low WRAT group' decision. In comparison, the high WRAT group (Fig. \ref{fig:gradcam_class1}) shows a relatively smaller number of significant FCs, which contributed to the 'high WRAT group' decision. This is further validated in the average histogram of the activation maps, i.e., Fig. \ref{fig:hist}. For the low WRAT group (Fig. \ref{fig:hist}-left), a large portion of FCs were activated (high grey-scale value), while for high WRAT group (Fig. \ref{fig:hist}-right), only a small portion of them were activated.

To identify significant brain FCs and SNPs, pixels with gray-value $> 0.05 \times$maximum gray-value were selected, following the instructions in the work \cite{selvaraju2017grad}. After that, FCs and SNPs with $>0.7$ occurring frequency across all subjects were further selected as significant FCs (see Figs. \ref{fig:fc_graph}-\ref{fig:fc_circle}) and SNPs (listed in Tables \ref{tab:gene_class0}-\ref{tab:gene_class1}).

The identified brain FCs (ROI-ROI connections) and their corresponding ROIs were visualized in Fig. \ref{fig:fc_circle} and Fig. \ref{fig:fc_graph}, respectively. For the high WRAT group (Fig.  \ref{fig:fc_circle}.b), three hub-ROIs (lingual gyrus, middle occipital gyrus, and inferior occipital gyrus) exhibited dominant ROI-ROI connections over the others. All of the three hub-ROIs are occipital-related. Lingual gyrus, also known as medial occipitotemporal gyrus, plays an important role in visual processing \cite{mechelli2000differential,mangun1998erp}, object recognition, and word processing \cite{mechelli2000differential}. The other two hubs, i.e., middle and inferior occipital gyrus, also play a role in object recognition \cite{grill2001lateral}. As shown in Fig. \ref{fig:fc_circle}.b, the hub-ROIs also connect to several other ROIs, e.g., cuneus, and parahippocampal gyrus. Among them, the cuneus receives visual signals and is involved in basic visual processing. The parahippocampal gyrus is related to encoding and recognition \cite{megevand2014seeing}. These suggest that the three occipital gyri are first activated when processing visual and word signals during the WRAT test, and then several downstream processing ROIs, e.g., para hippocampal gyrus, are activated for further complex encoding. As a result, strong FCs in these ROI-ROI connections may lead the gCAM-CCL to select the high WRAT group. 

For the low WRAT group (Fig.  \ref{fig:fc_circle}.a), there were no significant hub ROIs identified. Instead, several previously reported task-negative regions, e.g.,  temporal-parietal and cingulate gyrus \cite{hamilton2011default}, were identified. This indicates that the low WRAT group may be weaker in activating cognition-processing ROIs and therefore task-negative are relatively more active, which leads the gCAM-CCL to make the 'low WRAT group' decision.

As seen in Fig. \ref{fig:fc_circle}a-b, a relatively larger number of FCs contributed to the low WRAT group, compared to that of the high WRAT group. Despite this, as shown in Table \ref{tab:wrat_acc}, the sensitivity, however, is lower than the specificity, which means that the accuracy of classifying low WRAT group is lower. This suggests that the identified FCs for the high WRAT group are relatively more discriminative while the low WRAT group may contain more noisy FCs.

Gene enrichment analysis is conducted on the identified SNPs (Tables \ref{tab:gene_class0}-\ref{tab:gene_class1}) using ConsensusPathDB-human (CPDB) database\footnote{http://cpdb.molgen.mpg.de/}, and the enriched pathways are listed in Tables \ref{tab:pathway_class0}-\ref{tab:pathway_class1}. Several neurotransmission related pathways, e.g., regulation of neurotransmitter levels and synaptic signaling, are enriched from the identified high WRAT group genes. This suggests that the high WRAT group may have stronger neuron signaling ability. The stronger neuron-signalling may benefit the daily training and development of ROI-ROI connections, which may further contribute to stronger cognitive ability. For the low WRAT group, several brain development and neuron growth related pathways, e.g., midbrain development and growth cone, were enriched, which suggests that the low WRAT group may highlight problems in brain/neuron development. This may further affect the ROI-ROI connections, leading to weaker cognitive ability.

\section{Conclusion}
In this work, we develop an interpretable deep multimodal fusion model, namely gCAM-CCL, which can perform automated classification/diagnosis and result interpretation. The gCAM-CCL model can generate activation maps which indicate pixel-wise contribution of the inputs, e.g., images and genetic vectors, by first calculating each feature map's gradients and then merge the gradients using global average pooling to combine the feature maps. Moreover, the activation maps are class-specific, which further promotes class-difference analysis and biological mechanism analysis. 

The proposed model was applied to an imaging-genetic study to classify low/high WRAT groups. Experimental results demonstrate gCAM-CCL's superior performance in both classification and biological mechanism analysis. Based on the generated activation maps, a number of significant brain FCs and SNPs were identified. Among the significant FCs (ROI-ROI connections), three visual processing ROIs exhibited dominant ROI-ROI connections over the others. In addition, several signal encoding ROIs, e.g., the parahippocampa gyrus, showed connections to the three hub-ROIs. These suggest that during task-fMRI scans, object recognition related ROIs are first activated and then downstream ROIs get involved in further signal encoding. Results also suggest that high cognitive group may have higher neuron-transmitter signalling levels while low cognitive group may have problems in brain/neuron development, resulting from genetic-level differences. The results demonstrate that gCAM-CCL is superior in both classification and result interpretation, and therefore it can find wide applications in multimodal integration and imaging-genetic studies.

\section*{Acknowledgment}
The authors would like to thank the NIH (R01 GM109068, R01 MH104680, R01 MH107354, P20 GM103472, R01 EB020407, R01 EB006841, R01 MH121101, P20 R01 GM130447) and NSF (\#1539067) for the partial support.


\ifCLASSOPTIONcaptionsoff
  \newpage
\fi

\bibliography{dcl}
\bibliographystyle{IEEEtran}





\end{document}